\documentclass[prl,
twocolumn,
showpacs
]{revtex4-1}

\usepackage{amsmath}
\usepackage{amssymb}
\usepackage{amsbsy}
\usepackage{epsfig}
\usepackage{epstopdf}
\usepackage{graphicx}
\usepackage{times}
\usepackage{txfonts}
\usepackage{color}
\usepackage{dcolumn}
\usepackage[hidelinks]{hyperref}
\begin{document}

\title{Circular polarization memory in polydisperse scattering media}


\author{C.~M.~Macdonald}
\affiliation{Department of Physics, University of Otago, Dunedin, New Zealand}
\author{S.~L.~Jacques}
\affiliation{Departments of Biomedical Engineering and Dermatology, Oregon Health \& Science University, Portland, OR, USA}

\author{I.~V.~Meglinski}
\homepage[]{\\http://www.biophotonics.ac.nz}
\affiliation{The Dodd-Walls Centre for Photonic and Quantum Technologies, Department of Physics,
University of Otago, Dunedin, New Zealand \\
Opto-Electronics and Measurement Techniques Laboratory, University of Oulu, Oulu, FI-9014, Finland \\
Interdisciplinary Laboratory of Biophotonics, Tomsk State University, Tomsk 634050, Russia}

%\thanks{}
%\affiliation{}


\date{\today}

\begin{abstract}
We investigate the survival of circularly polarized light in random scattering media. The surprising persistence of this form of polarization has a known dependence on the size and refractive index of scattering particles, however a general description regarding polydisperse media is lacking. Through analysis of Mie theory, we present a means of calculating the magnitude of circular polarization memory in complex media, with total generality in the distribution of particle sizes and refractive indices. Quantification of this memory effect enables an alternate pathway towards recovering particle size distribution, based on measurements of diffusing circularly polarized light.
\end{abstract}
%\keywords{}
\maketitle
\section{I. INTRODUCTION}
An understanding of scattered radiation is deeply embedded within such scientific disciplines as astronomy, meteorology, climatology, and more recently biomedical imaging~\cite{Boas}. Measurement of this radiation allows for the remote detection of object properties, through inspection of both the spectral, and spatial redistribution of energy. Notably, the inclusion of polarization in this analysis serves to enrich the source of available information. An area of immense interest concerning such behavior, is the study of optical radiation scattering within biological media~\cite{Vitkin}. It was first realized that through appropriate filtering, polarized light could enable the selective imaging of either surface, or subsurface tissue layers~\cite{Schmitt}. Since then, similar applications have helped to improve imaging modalities such as optical coherence tomography~\cite{De Boer}, and light scattering spectroscopy~\cite{Gurjar}. In addition to the advancement of existing techniques, the recovery of physical properties solely based on the observed transformation of polarized light is the objective of much research~\cite{Vitkin}. Interestingly, in media consisting of particles comparable to the wavelength in size (Mie regime), circular polarization can survive many more scattering events than linear polarization due to excessive forward scattering. Linear states depolarize once the direction of the incident photon stream is eventually isotropized~\cite{XuAlfano2}. On the other hand, due to the preference of shallow angle scattering which generally preserves helicity, incident circularly or elliptically polarized states persist on a longer length scale. This effect is known as polarization memory \cite{MacKintosh,XuAlfano,KimOL}. The extent of the polarization memory is strongly dependent on the size parameter~\cite{Bicout,Cai}, and on the refractive index mismatch between the particles and background medium~\cite{KimMoscoso1}. This offers an additional gauge by which to characterize medium properties, as measurements of the polarization state of multiply scattered light contain this information, even beyond the distance required to randomize photon direction, known as the transport length $l_{t}$.

In this article, we address the dynamic relationship between circular polarization memory and the particles present in the scattering medium. Recent studies have shown that the survival of linear polarization exceeds that of circular polarization in many biological tissues~\cite{Alali_Vitkin,Ahmad_Vitkin,Sankaran1}. Such behavior is characteristic of smaller Rayleigh scattering particles, although these tissues exhibit anisotropic scattering typical of the Mie regime. This has suggested an inability to predict depolarization rates in complex media based purely on bulk scattering properties such as the reduced scattering coefficient $\mu_{s}'$, the absorption coefficient $\mu_{a}$, and the anisotropy $g$ factor~\cite{Gosh2,Antonelli}.
Here, we demonstrate a quantitative description of circular polarization memory in media consisting of a variety of independent scattering spheres, with total generality in particle size, and refractive index distribution. Our method uses direct analysis of the single scattering behavior attained from Mie theory, and a flexible integral approach towards characterizing the depolarization rate within complex media. This will allow for additional information to be gathered from diffusing light measurements, and clarify the connection between particle distribution, and the observed crossover between scattering regimes.
\section{II. HELICITY SURVIVAL PARAMETER}
\begin{figure}[ht]
 \begin{center}
  \epsfxsize=8.5cm
  \epsfbox{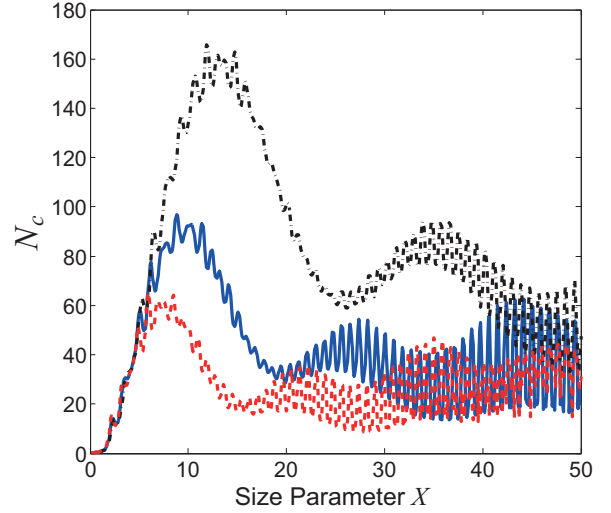}
  \caption{(Color online). Helicity survival parameter $N_{c}$ of perfectly monodisperse systems of scattering spheres with size parameter $X$, and refractive index ratio $m = 1.15$ (dash-dotted black), $m = 1.20$ (solid blue), and $m = 1.25$ (dashed red).  }
 \label{Depolarization}
 \end{center}
\end{figure}

Mie theory offers an exact solution for scattering of a plane wave by a single spherical particle of any size. The relationship between the incoming, and scattered Stokes vectors describing polarization can be expressed in terms of the scattering matrix~\cite{Bohren}:
\begin{equation}
\begin{bmatrix}
I_{s}\\
Q_{s}\\
U_{s}\\
V_{s}
\end{bmatrix}
=\begin{bmatrix}
S_{11}(\theta) &S_{12}(\theta)  &0  &0 \\
S_{12}(\theta) &S_{11}(\theta)  &0  &0 \\
 0&  0& S_{33}(\theta)& S_{34}(\theta) \\
 0&  0&  -S_{34}(\theta)& S_{33}(\theta)
\end{bmatrix}
\begin{bmatrix}
I_{i}\\
Q_{i}\\
U_{i}\\
V_{i}
\end{bmatrix} \;,
\end{equation}
where the elements of the scattering matrix are dependent on the scattering angle $\theta$, and the Stokes vectors are both defined with an orientation aligned to the scattering plane. For an incident right-circularly polarized state, with $I_{i} = V_{i}  =1$, $Q_{i} = U_{i} = 0$, the Stokes vector for the partial wave scattered in the $\theta$ direction, with $\phi$ azimuth is:
\begin{equation}
\begin{bmatrix}
I_{s}\\
Q_{s}\\
U_{s}\\
V_{s}
\end{bmatrix}
=
\begin{bmatrix}
S_{11}(\theta)\\
S_{12}(\theta)\\
S_{34}(\theta)\\
S_{33}(\theta)
\end{bmatrix} \;,
\end{equation}
which is identical for all $\phi$ due to the rotational invariance of the input Stokes vector. The expression for the degree of circular polarization of this partial wave is:
\begin{equation}
\frac{V_{s}}{I_{s}} = \frac{S_{33}(\theta)}{S_{11}(\theta)}\;,
\end{equation}
while the ensemble average of this quantity, denoted $P_{c}$, over all scattering angles, weighted by the intensity of each partial wave is thus,
\begin{equation}
P_{c} = \frac{\int_{0}^{2\pi} \frac{S_{33}(\theta)}{S_{11}(\theta)}S_{11}(\theta) \;d\theta}{\int_{0}^{2\pi}S_{11}(\theta) \; d\theta} \;.
\label{integral}
\end{equation}
After any arbitrary number of scattering events $n$ from identical independent spheres, the value of $P_{c}$ for all partial waves can be shown to follow an exponential decay~\cite{XuAlfano}. Here we will define the characteristic helicity survival parameter $N_{c}$, which is a measure of the number of scattering events required to depolarize an incident circular state. The average degree of circular polarization as a function of scattering order $n$ is then:
\begin{equation}
P_{c}(n) = e^{-n/N_{c}} \; \;,
\label{exponential}
\end{equation}
where $N_{c}$ can be found from the single scattering behavior calculated in Eq.~(\ref{integral}) and the natural logarithm of Eq.~(\ref{exponential}) with $n = 1$. This procedure can be carried out for a monodisperse system for any size parameter $X = 2\pi a n_{b}/ \lambda$, and any refractive index ratio $m = n_{s}/n_{b}$, where $n_{s}$ and $n_{b}$ are the refractive indices of the sphere, and background respectively, and $a$ is the particle radius. Figure~\ref{Depolarization} shows the calculated survival parameter for a range of sphere sizes, and refractive indices. The value of $N_{c}$ can be seen to fluctuate on both large and small scales with respect to a change in the size parameter $X$.
\section{III. POLYDISPERSE MEDIA}
For a medium consisting of a range of particle sizes, Eq.~(\ref{exponential}) must be generalized to a distribution of exponentials:
\begin{equation}
P_{c}(n) = \int_{0}^{\infty}I_{f}(X)e^{-n/N_{c}(X)}\;dX \;,
\label{Integral}
\end{equation}
where $I_{f}(X)$ is appropriately normalized such that,
\begin{equation}
\int_{0}^{\infty} I_{f}(X)\;dX = 1\;.
\end{equation}
$I_{f}(X)$ is the fraction of total intensity scattered, on average, by particles of size $X$ with corresponding helicity survival parameters given by $N_{c}(X)$. This function is directly proportional to the scattering coefficient of each particle size, $I_{f}(X) \propto \mu_{s}(X)$. In other words, the resulting circular depolarization rate of a polydisperse system is dependent on not only the helicity survival parameters of the constituent particles, but also the relative scattering cross sections $\sigma_{s}(X)$, and number densities $\rho_(X)$. We then have:
\begin{equation}
I_{f}(X) = \frac{\rho(X)\sigma_{s}(X)}{\int_{0}^{\infty}{\rho(X)\sigma_{s}(X)\;dX}} \;,
\end{equation}
and finally, in terms of the volume fraction distribution $f(X)$:
\begin{equation}
I_{f}(X) = \frac{\frac{f(X)}{\nu(X)}\sigma_{s}(X)}{\int_{0}^{\infty}{\frac{f(X)}{\nu(X)}\sigma_{s}(X)\;dX}} \;,
\end{equation}
where $\nu(X)$ is the single particle volume for spheres of size parameter $X$. Equation~(\ref{Integral}) integrates over the helicity survival parameters of all particles present in the medium, with each $N_c$ being calculated from the single scattering behavior. Thus, with a database of values stored for $N_{c}(X)$, and $\sigma_{s}(X)$, the effective helicity survival parameter for any polydisperse system of spheres defined by the distribution function $I_{f}(X)$ can be found immediately, a key advantage of this approach.

\begin{figure}[t]
 \begin{center}
  \epsfxsize=8.5cm
  \epsfbox{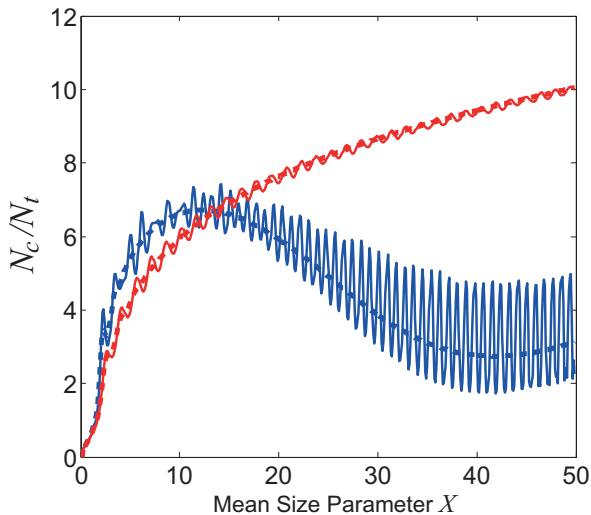}
  \caption{(Color online). Circular polarization memory ($N_{c}/N_{t}$) for a range of size parameters, refractive index ratio $m = 1.04$ (red upper), and $m = 1.2$ (blue lower). Monodisperse media (solid lines), and polydisperse media with a $10\%$ Coefficient of Variance in particle size (dashed lines).}
 \label{Poly}
 \end{center}
\end{figure}

Narrow polydispersion in scattering media effectively washes out the short scale fluctuations of particle size on the helicity survival $N_{c}$. For example, media with particle number densities $\rho(X)$ obeying a normal distribution with a small ($10\%$) coefficient of variance in particle size are shown in Fig.~\ref{Poly}. Here the ratio $N_{c}/N_{t}$ is used to show the extent of the circular polarization memory, where $N_{t}$ is the number of scattering events required to randomize photon direction \textit{i.e.} the number of scattering events in one transport length $l_{t}$. Narrow polydisperse distributions shown in Fig.~\ref{Poly} are common in media such as solutions of polystyrene microspheres in water ($m = 1.2$), where the coefficient of variance is typically $2-10\%$. These solutions are widely used as simulating media, or ``phantoms", due to their well defined bulk scattering properties.

\begin{figure}[b]
 \begin{center}
  \epsfxsize=8.5cm
  \epsfbox{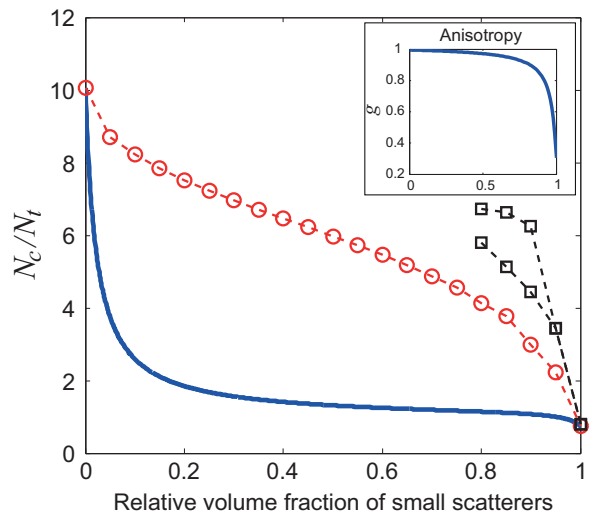}
  \caption{(Color online). (solid line) - Circular polarization memory ($N_{c}/N_{t}$) verses relative volume fraction of small scatterers of size $X = 1.3 \pm 10\%$, with remaining scattering material composed of particles with $X = 50 \pm 10\%$. Refractive index ratio is $m = 1.04$. ``Monodisperse" phantoms with equivalent macroscopic scattering anisotropy, refractive index ratio $m = 1.040$ (circles), and  $m = 1.2$ (squares). (Inset) - Matched scattering anisotropy of all three types of media.}
 \label{Cell}
 \end{center}
\end{figure}

\begin{figure}[t]
 \begin{center}
  \epsfxsize=8.5cm
  \epsfbox{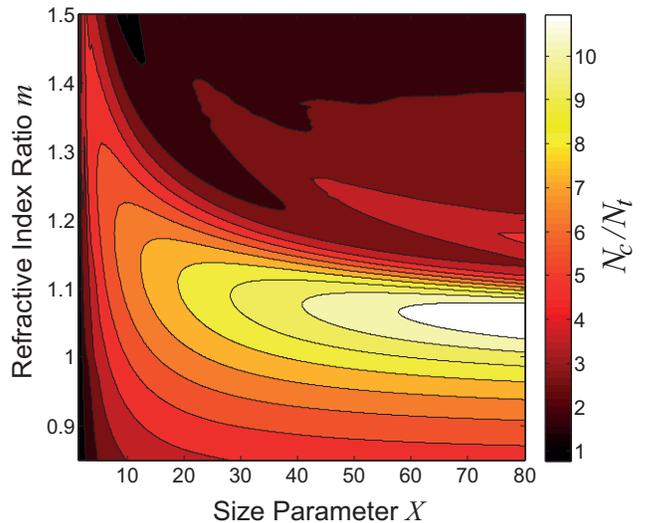}
  \caption{(Color online). Contour plot of circular polarization memory ($N_{c}/N_{t}$) for size parameter $0.1 < X < 80$, and refractive index ratio $ 0.85 < m < 1.5$. Each point represents the circular polarization memory for a narrow polydispersion ($10\%$ coefficient of variance) about the mean size parameter $X$.}
 \label{Heatmap}
 \end{center}
\end{figure}

Scattering from more complex materials such as biological tissue can include contributions from a wide array of particle sizes. Disagreement between measurements of circularly polarized light scattered from tissue samples, and from matched phantoms has been reported~\cite{Ahmad_Vitkin,Sankaran2}. This is likely due in part to the presence of both large Mie scattering structures in tissue such as cell nuclei (typically $5-10$ $\mu$m), as well as smaller Rayleigh-Gans scatterers (mitochondira, lysosomes \textit{etc.} $<0.5$ $\mu$m)~\cite{Schmitt2}. These phantom studies often use a monodisperse solution with particle size and concentration chosen to match the reduced scattering coefficient $\mu_{s}' = \mu_{s}(1-g)$, and the scattering anisotropy $g$, to another more complex medium. However this is an oversimplification, and fails to reproduce the depolarization behavior, as we shall demonstrate. To approximate scattering from cell nuclei with smaller interspersed cellular organelles, we have calculated $N_{c}$ for media where the volume fraction contributions from small scatterers, and large scatterers have been varied. Here, $f(X)$ is chosen to take a bimodal form, with one normally distributed contribution with size parameter $X = 1.3 \pm10\%$, and another normally distributed contribution at $X = 50\pm 10\%$. The fraction of the total volume of scattering material taken by the small scatterers is varied from $0$ (nuclei only scattering) to $1$ (small organelle only scattering). The refractive index ratio used is $m = 1.04$ which is typical of biological media with $n_{s} \approx 1.42$ (organelles), and $n_{b} \approx 1.37$ (cytoplasm)~\cite{Schmitt2}. The results are shown in Fig.~\ref{Cell}.
It can be seen that the circular polarization memory rapidly deteriorates with the addition of small scatterers to the medium. For relative volume fractions of small particles above $10\%$, the effectiveness of the medium to preserve multiply scattered circularly polarized light is significantly diminished. To emphasize the importance of adequately accounting for polydispersion in scattering media, we have also calculated the circular polarization memory ($N_{c}/N_{t}$) for a set of media with only one dominant size. These points on the graph~(Fig.~\ref{Cell}) represent typical ``monodisperse" optical phantoms with a $10\%$ size distribution about the mean diameter. In each case, these phantoms have been chosen to match the scattering anisotropy of the bimodal distribution used to model tissue. The circles represent phantoms consisting of particles with $m = 1.04$. The squares are calculated with $m = 1.2$, to represent polystyrene phantoms. The branching of these points is due to the multiple possibilities in particle size for certain anisotropy $g$ values, and the partial absence of data points expresses that the upper limit of scattering anisotropy for polystyrene microspheres is $g = 0.93$ (in water). It is clear from the figure that a significant overestimation is made of circular polarization memory when such phantoms are used, even if they are chosen such that $\mu_{s}$, $g$, and hence $\mu_{s}'$ are equivalent. The shortcomings of monodisperse phantoms evident here are in direct agreement with the previously mentioned studies~\cite{Ahmad_Vitkin,Sankaran2,Antonelli}. It is important to note that the ability of small scatterers to reduce the polarization memory of larger constituent particles is heavily dependent on the relative scattering cross section $\sigma(X)$. In the example situation shown in Fig.~\ref{Cell}, $X = 1.3$ spheres were chosen because they are more effective at reducing circular polarization memory than smaller ($X<1.1$) spheres, which have intrinsically lower values of $N_{c}$. This is due to the trade off between faster depolarization, but lower scattering efficiency as size is reduced. This demonstrates that depolarization of incident circular states within polydisperse media is an intricate process dependent on the specific size and refractive indices of each contributing particle. Equation~(\ref{Integral}) can be generalized further to include variations in particle refractive index mismatch $m$:
\begin{equation}
P_{c}(n) = \int_{0}^{\infty}\int_{0}^{\infty}I_{f}(X,m)e^{-n/N_{c}(X,m)} \;dm\;dX \;,
\label{General}
\end{equation}
where $I_{f}(X,m)$ now represents the fraction of total intensity scattered by particles for which $N_{c} = N_{c}(X,m)$. The effective helicity survival parameter of the polydisperse medium can once again be found by inverting Eq.~(\ref{exponential}) after evaluating this integral. To further highlight the flexibility of the approach presented here, Fig.~\ref{Heatmap} shows the magnitude of circular polarization memory for a wide range of size parameters $X$, and refractive index ratios $m$. Each point on the image represents $N_{c}/N_{t}$ for a $10\%$ particle size distribution about the mean size parameter $X$. Incident circularly polarized light is seen to be preserved most in media where both $\left \vert m-1 \right \vert << 1$, and particle size is large. Such properties are consistent with large cellular structures in tissue. A complex cellular model could be proposed by the integration over all types of organelles present, including vacuoles with $m<1$.  However, the closely packed nature of scattering centers in many biological tissues may cause discrepancies from such models when dependent scattering arises, as shown in Ref.~\cite{Sankaran3}. Thus, care is required when generalizing these findings which are a result of the independent scattering approximation for spherical particles.
\section{IV. SUMMARY}
In summary, we have presented here a method for characterizing circular polarization memory in polydisperse media, where the particle distribution has been demonstrated to intricately determine the decay rate of incident circularly polarized light: Narrow polydispersion about one dominant size parameter has the effect of smoothing out the short scale fluctuations seen with theoretical solutions containing perfectly identical centers. This type of polydispersion is typical of phantom media such as polystyrene microspheres, and demonstrates that models using pure monodispersion can differ somewhat from realistic media where statistical variations in particle size are unavoidable (see Fig.~\ref{Poly}). In complex media, where size distribution spans from the Rayleigh regime to the Mie regime, the extent of circular polarization memory is shown to be greatly impacted by only a small volume fraction of sub-wavelength scatterers. In our particular example (Fig.~\ref{Cell}) we showed that circular polarization decays within approximately two transport lengths ($N_{c}/N_{t} < 2$) for media consisting of over $20\%$ relative volume fraction of small $1.1 < X < 1.5$ particles. With the method put forward here, it is now possible to rapidly calculate the helicity survival parameter $N_{c}$ for a medium with any general form of polydispersion in both size parameter $X$, and refractive index ratio $m$. Thus, recovery of this parameter from degree of polarization measurements of semi-diffuse light can provide an alternate means towards characterizing particle distribution. Furthermore, although circular polarization memory
is a manifestation of anisotropic scattering, we have demonstrated the important distinction that it cannot be characterized solely by the scattering anisotropy $g$ parameter.

%%%%%%%%%%%%%%%%%%%%%%%%%%%%

We acknowledge support from the William Evans fellowship provided by the University of Otago for Professor Steven Jacques, The Dodd-Walls Center for Photonic and Quantum Technologies, and the Department of Physics at the University of Otago.
%%%%%%%%%%%%%%%%%%%%%%%%%%%%
%%%%%%%%%%%%%%%%%%%%%%%%%%%%

\end{document}